# Engineering MoS$_2$-MoTe$_2$ Heterojunctions: Enhancing Piezoresponse and Rectification

Sai Saraswathi Yarajena[1], Akshay K. Naik[2]

1: saiyaras.95@gmail.com/sy785@cornell.edu , 2: anaik@iisc.ac.in

*Centre for Nano Science and Engineering, Indian Institute of Science, Bangalore-560012, India.*

## Abstract:

Piezoelectric materials play a vital role in energy harvesting, piezotronics and various self-powered sensing applications. The piezoelectric strength of 2D materials is limited by the carrier charge screening, leading to reduced open circuit voltages and poor piezotronic performances. Reducing the carrier screening in devices is a key requirement to fully utilize the potential of 2D materials for piezoelectric applications. In this work, we demonstrate that lateral heterojunction devices offer an excellent way to improve the piezoelectric open circuit voltages and rectification ratios. Because of the asymmetric contacts with Nickel (Ni) electrodes, the heterojunctions of monolayer(1L) MoS$_2$ and MoTe$_2$ form a hybrid Schottky/p-n diode. We demonstrate a rectification ratio of more than 5000 without electrostatic gating. We observed that devices with higher junction potentials exhibit piezoelectric open-circuit voltages exceeding 1V and a peak power density of 690 mW/m² The output characteristics reveal a trade-off between open circuit voltages and rectification ratios. These findings and the role of built-in (cut-in) voltages in energy harvesting provide valuable insights for the design of piezotronic junctions to achieve high piezoelectric output and/or rectification ratios. Design aspects of heterojunctions discussed in this manuscript can be applied to other emerging nanomaterials.

## Introduction

In the field of energy harvesting, piezoelectric devices and triboelectric nanogenerators are extremely important[1–3]. In addition to energy harvesting, piezoelectric semiconductor materials have electronic properties that enable integrated energy harvesting and sensor systems[4]. With their exotic electronic properties, 2D materials are touted as the next-generation materials for electronic devices[5–7]. Moreover, the piezoelectricity property in some of these 2D materials makes them an excellent choice for piezotronics[8]. Among the 2D materials, transition metal dichalcogenides (TMDCs) such as MoS$_2$ and MoTe$_2$ are well known for their stability in the air, and the growth of these materials is well studied[9–11]. Theoretical studies estimate an open circuit voltage of about 350 mV for a monolayer MoS$_2$ device[12]. However, practical piezotronics devices with only MoS$_2$ generate open circuit voltages in the range of a few millivolts(~15 mV)[13]. The low piezoelectric output in these devices is the result of charge screening, large electrode capacitance and fermi-level pinning[13]. This low piezoelectric output will limit the functionality of these devices fabricated with a single 2D material.

A rectifying heterojunction formed with piezoelectric 2D materials can improve the piezoelectric output by reducing the charge screening effects at the junctions. Furthermore, the rectifying response observed in these devices is useful for many applications, such as solar cells, switches and photodiodes [14]. However, very few reports have explored 2D material heterojunctions for piezotronics[15–17]. Wu et al. [18] proposed a p-n junction with MoS$_2$ and WSe$_2$ for piezophototronic applications. The ON-OFF current ratio in these devices was limited to about 200 at ±1V. The reported open circuit potentials did not indicate any significant advantage of heterojunction devices over single MoS$_2$ devices reported in the literature[16,18,19]. ZnO nanowire heterojunctions were studied to reduce the charge screening and enhance the piezoelectric outputs[20]. However, the enhancement in piezoelectric response is not substantially high even with these ZnO nanowire heterojunctions. Furthermore, the junctions play a key role in charge carrier dynamics and energy conversion. However, to the best of our knowledge, reports in the literature have not looked at the effect of junction characteristics on the piezoelectric signal or the role of cut-in voltages of the rectifying junction on the piezotronic properties.

In this work, we study and discuss the properties of MoTe$_2$ and MoS$_2$ lateral heterojunction devices fabricated on flexible substrates and correlate the change in junction characteristics with the piezoelectric effect. Additionally, we emphasize the prerequisites to reduce the charge screening effects. The results of strain-dependent behaviour in several monolayer MoS$_2$ (n-type) and MoTe$_2$ (p-type) heterojunctions unveil interesting piezotronic properties that have the potential to achieve high piezoelectric output or high rectification. The trade-off between rectification ratio and piezoelectric output with cut-in voltages is a critical aspect for the design of high-performance piezotronic devices tailored to specific requirements.

## Results and discussion

We have selected a junction of 1L-MoS$_2$ and a few layers of MoTe$_2$ for our devices. Mechanically exfoliated MoS$_2$ and MoTe$_2$ flakes are used to make the devices on Nano flex substrates[21] with sputtered silicon dioxide (SiO$_2$) as a base layer (fig. 1(a)). The choice of these material and substrates was based on many different requirements. For good piezoelectric response, low screening and adhesion of 2D materials are essential. Free carriers and charged impurities introduced in a material during the processing will recombine with the charges generated due to the piezoelectricity. This results in a net reduction of polarization. This phenomenon is called internal screening[1]. In 2D materials, internal screening can occur due to defect states such as vacancies, dopant atoms and charged impurities. Internal screening can be reduced by suppressing the free carrier density and charged impurities. Impurity charge screening is the dominant screening effect in two-dimensional TMDCs[22]. A junction of p and n-type semiconductors or the Schottky junctions utilizes these free carriers to form a depletion region, thus reducing the total amount of screening.

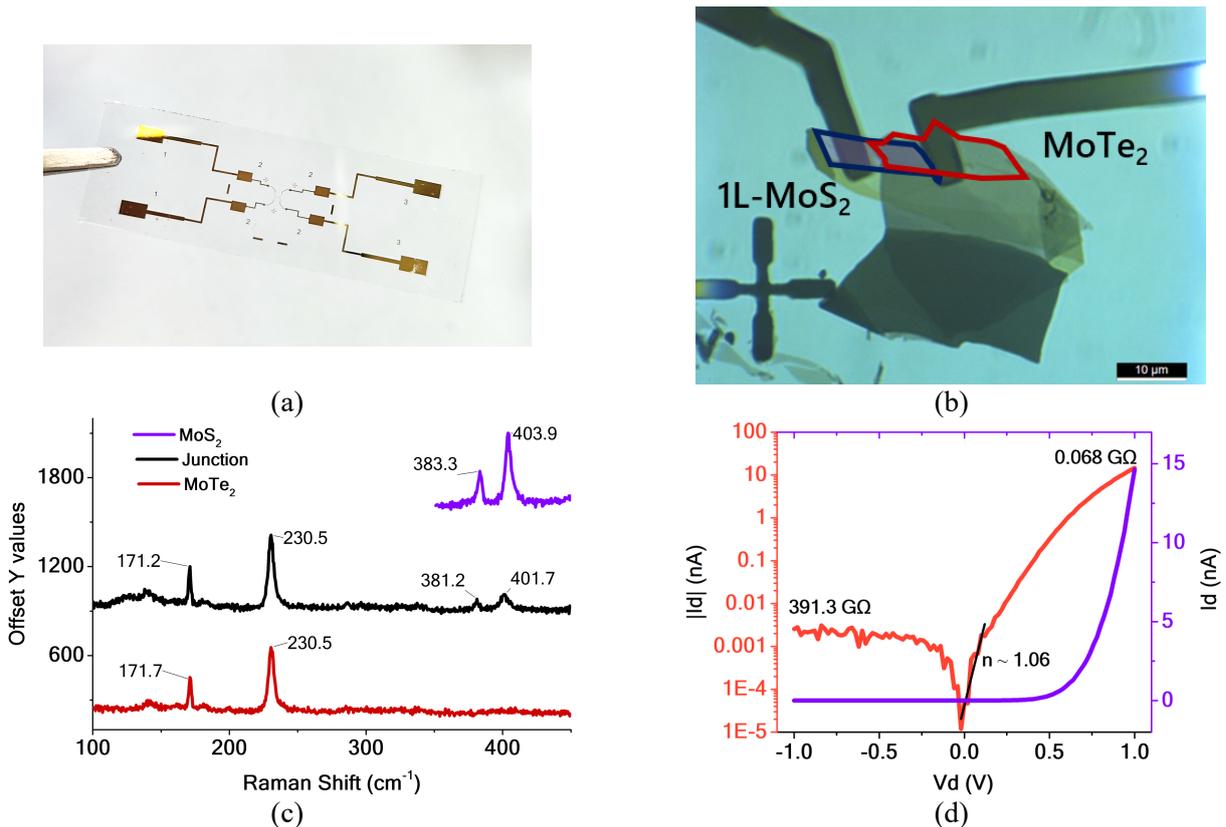

*Figure 1: (a) Image of fabricated heterojunction device on flexible substrate, (b) Optical micrograph of heterojunction device with MoS$_2$(outlined in blue) and MoTe$_2$(outlined in red) junction (spatial polariser is used to improve the optical images), (c) Raman spectra on MoS$_2$, MoTe$_2$ and their junction region, (d) I-V characteristics of heterojunction device in linear and log scales. Resistance values at ±1 V and the ideality factor (n) are mentioned in the graph.*

A corresponding effect that has its origin at the material's interface is called external screening. This is often the result of dangling bonds at the metal and piezoelectric material interface. Various chemical species from the environment and those generated during the device fabrication process adsorb at the interface and promote the screening of charges [1]. The rate of external screening also depends on the type of metal contact used. Schottky metal contacts are preferred for piezoelectric generators instead of ohmic contacts[1]. With thermionic emission, mobility and carrier density are high in ohmic contacts, resulting in faster screening. On the other hand, Schottky contacts have a depletion region. This depletion region adds capacitance at the junction, thereby reducing the rate of screening to improve piezoelectric output. For example, SnSe has the highest reported piezoelectric output in the literature, with an open circuit voltage of 0.7 V[23]. However, SnSe is also affected by high carrier screening because of its low bandgap and the expected ohmic contact with metal electrodes. We have also observed similar screening effects in $MoTe_2$ devices with metal contacts (Ni/Au) fabricated on flexible substrates (see SI). Although $MoTe_2$ is reported to have the highest piezoelectric coupling of all TMDCs[8], a few-layer $MoTe_2$ alone is not an efficient piezotronic material for direct piezoelectricity applications. However, $MoTe_2$ has a relatively low bandgap of ~0.9-1 eV and low resistance (in MΩ), and high ON currents can be achieved with its junctions. Hence, it is chosen as one of the materials for heterojunction. In contrast, monolayer $MoS_2$ boasts a comparatively substantial bandgap of 1.89 eV. It also showcases remarkable resistance, measuring in the tens of GΩ, hence making it the preferred second material. The combination enables a resistance switching from very low to very high, i.e. a high rectification ratio.

2D materials exfoliated onto polymer substrates have poor adhesion and slip during strain measurements[24]. This can be avoided by depositing a silicon dioxide ($SiO_2$) layer onto the polymer before exfoliating 2D materials. Furthermore, the inorganic dielectric layer $SiO_2$ helps reduce charge screening effects[21]. Hence, all our 2D devices on flexible substrates have a sputtered $SiO_2$ layer, which is deposited using magnetron sputtering at room temperature.

With this choice of materials, a lateral heterojunction device is formed with one metal contact on $MoS_2$ (outlined in blue) and the other on $MoTe_2$ (outlined in red), as shown in fig. 1(b). Raman spectroscopy is carried out on the heterojunction device, and fig. 1(c) shows the Raman data on 1L-$MoS_2$, few layer $MoTe_2$ and their junction region. The few-layer $MoTe_2$ has an $A_{1g}$ peak at 171.2 $cm^{-1}$ and $E^1_{2g}$ at 230.5 $cm^{-1}$. From this, the thickness of $MoTe_2$ is estimated to be more than five layers. The $A_{1g}$ and $E^1_{2g}$ peaks of $MoS_2$ have shifted towards lower frequencies on the junction region, but the difference between the peak positions has remained the same. The shift in the Raman peaks to lower frequencies is due to the tensile strain on the material. If in-plane stress was applied, only the $E^1_{2g}$ peak shift would shift left, and $A_{1g}$ would remain at the same position[25]. But here, both the peaks are shifted left, which could mean the presence of in-plane and out-of-the-plane tensile strain for 1L-$MoS_2$ on $MoTe_2$. The observed red shift in the $MoS_2$ Raman peaks on the junction is likely attributed to interlayer coupling between $MoS_2$ and $MoTe_2$ at the junction[26].

I-V characteristics of the device in linear and log scales are shown in fig. 1(d). The resistance changed from ~390 GΩ at –1 V to 68 MΩ at +1 V. This device has a rectification ratio of more than 5000, which is an order of magnitude higher than the other 2D material based junctions reported in the literature. Note that this rectification is achieved without electrostatic gating. The forward bias cut-in voltage is 0.15 V (the cut-in voltage is calculated as the voltage bias at which the current response changes from linear to exponential form[27]). An excellent ideality factor of 1.06 is obtained for this device (see SI ). The device with an ideality factor close to 1 is considered an ideal rectifying junction. In practice, the value of the ideality factor is between 1 and 2. A low ideality factor in our device indicates the absence of interface traps and other recombination mechanisms[14].

$MoS_2$ is an n-type semiconductor, and $MoTe_2$ is a p-type in ambient conditions, which results in a pn junction at their interface. The Ni metal electrode forms a Schottky contact with (1L) $MoS_2$ and an ohmic contact with the few-layer $MoTe_2$. Thus, there is a Schottky diode at the Ni/Au-$MoS_2$ region and a pn junction at the $MoS_2$-$MoTe_2$ region (fig. 2(a)). We call it a Schottky-pn hybrid junction. This is unlike $MoS_2$-$WS_2$ junctions reported in the literature, which had Schottky contacts at both ends[18]. The flat band diagram of $MoS_2$-$MoTe_2$ is shown in fig. 2(b). The net cut-in voltage ($V_{cut-in}$) to drive the device in forward bias is the built-in potential at the pn junction and the $MoS_2$-Ni/Au junction barrier. In forward bias (Vd > 0), a positive voltage is applied to $MoTe_2$ and $MoS_2$ terminal is grounded. The corresponding energy band diagram in forward bias is shown in fig. 2(c). The net barrier height reduces with applied forward bias, as shown in fig. 2(c), and the device starts conducting with high ON currents. The barrier height increases with reverse bias (Vd < 0 V), as shown in fig. 2(d), resulting in large resistance. The extent of resistance change under forward and reverse bias conditions after the cut-in voltage decides the rectifying ratio.

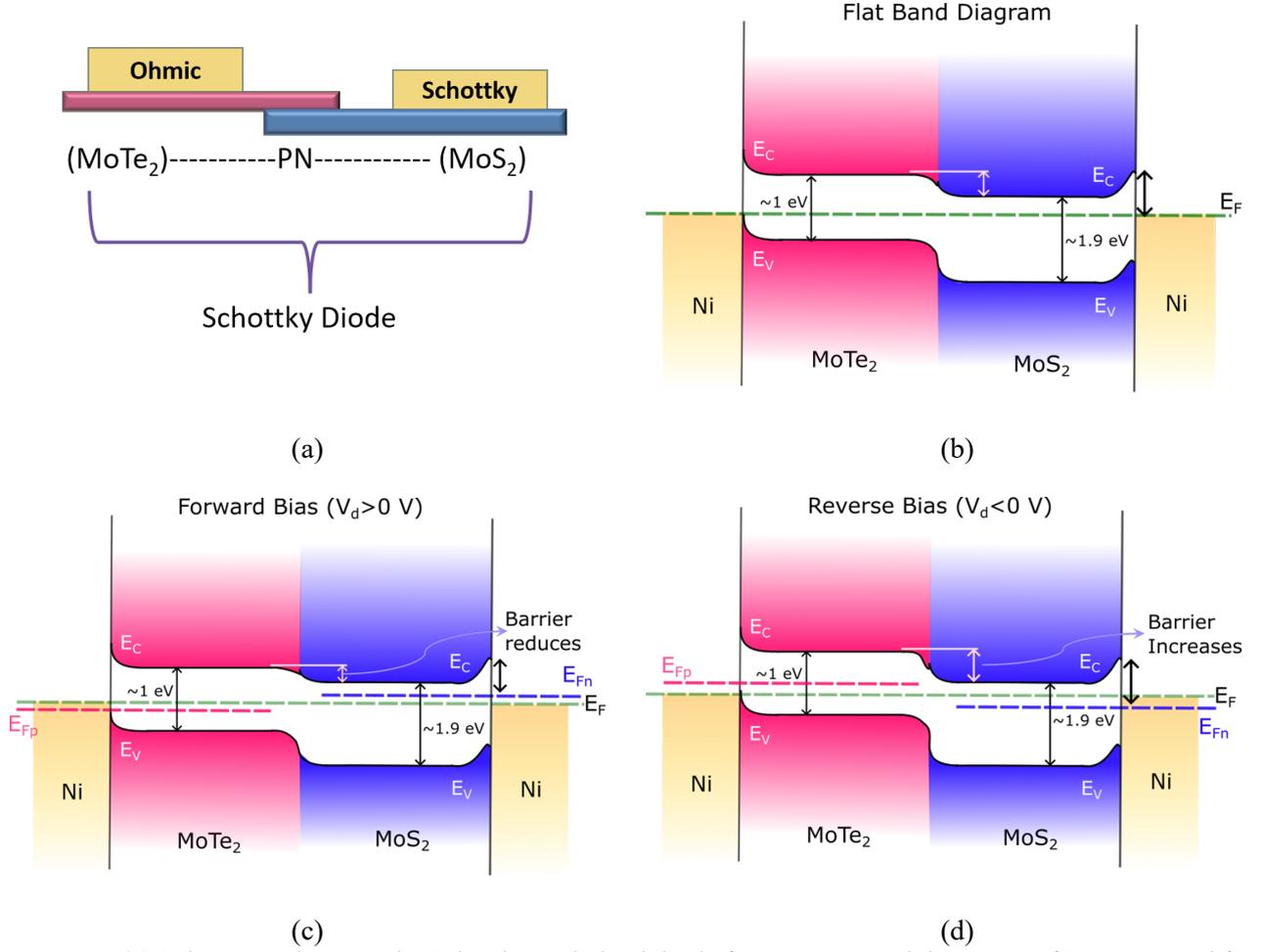

*Figure 2*: (a) Schematic showing the Schottky-pn hybrid diode formation, Band diagrams of 1L-MoS$_2$ and few-layer MoTe$_2$ device with Ni contacts in (b) flat band condition and (c) under forward bias ($V_d$ > 0 V), (d) under Reverse Bias ($V_d$ < 0 V)(Note: Band diagrams are only for demonstration purposes and not to be scaled).

We have fabricated multiple heterojunctions with 1L-MoS$_2$ and multilayer MoTe$_2$. The optical images of three of the devices (D1, D2, D3) with different rectification ratios (Table 1) are shown in fig. 3(a-c). The stacking of these exfoliated flakes with heterojunctions could not be controlled precisely with our current setup, and the yield of working devices was low. The junction area and the areas of 1L-MoS$_2$ and MoTe$_2$ varied across devices, resulting in differences in their ON-OFF resistances and cut-in voltages.

The rectification ratio primarily depends on the cut-in voltage for the device, off resistance, and ideality factor. It also depends on device geometry and interface trap states. The interface trap states are unpredictable and difficult to control in 2D material devices. The MoTe$_2$ layer should be thick, and MoS$_2$ should be thin, i.e., a monolayer, to achieve a high rectification ratio in a heterojunction. This ensures high ON and low off currents. Although MoS$_2$ is monolayer in all these devices, the metal electrode is contacting the relatively thick layers attached to monolayer MoS$_2$. This decreases the OFF resistance. Nonetheless, (1L) MoS$_2$-MoTe$_2$ devices could achieve rectification ratios that are much higher than those reported in the literature[15,16,28].

**Table 1**: List of three devices with variable rectification ratio

| Device Name | Rectification Ratio | Cut-in/Threshold Voltage (V) |
|---|---|---|
| D1 | 5725 (High) | 0.26 (Low) |
| D2 | 53 (Low) | 0.92 (High) |
| D3 | 3752 (Moderate) | 0.52 (Moderate) |

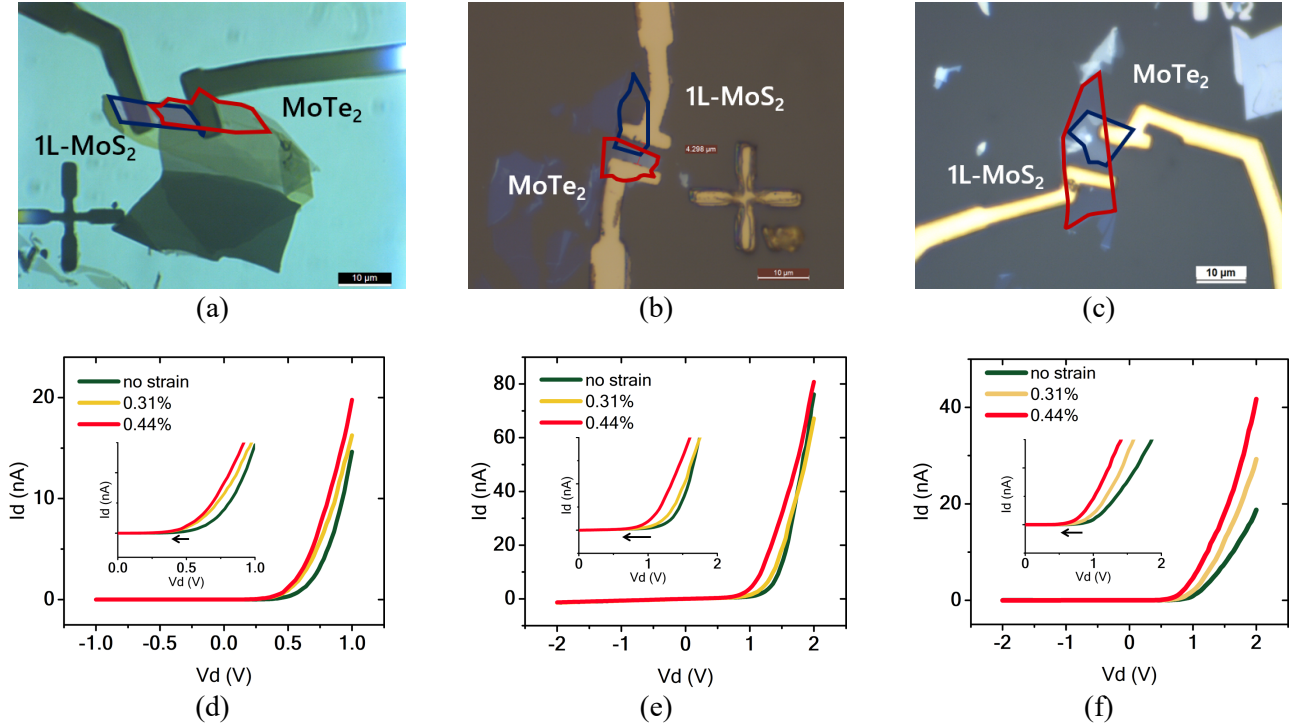

***Figure 3***: (a-c) Optical micrographs of devices D1, D2 and D3. Strain-dependent I-V characteristics of the devices (d) D1, (e) D2, and (f) D3. The change in the cut-in voltage ($V_{cut-in}$) from nominally zero strain to 0.44% strain is labelled on the graph.

Junction properties of these heterojunctions possibly depend on the MoTe$_2$ thickness, relative areas of MoS$_2$, MoTe$_2$ and the junction, and interface traps. Fig. 3(d-f) shows the strain-dependent I-V characteristics of these three devices. The change in cut-in voltage with the applied tensile strain of the three devices (D1, D2, D3) is indicated with an arrow in the plots shown in fig. 3(d-f). With an increase in the tensile strain, $V_{cut-in}$ of all the devices reduces. The applied tensile strain reduces the bandgap of the material and induces piezoelectric polarisation in monolayer MoS$_2$, thereby reducing the Schottky barrier height (fig. 4a). The reduction in barrier height is likely the reason for the decrease in $V_{cut-in}$. Fig. 4(b) shows the change in the rectification ratios with applied tensile strain (0.44%) in devices D1, D2 and D3. The rectification ratios of these devices also change with applied strain, indicating that the junction properties are indeed tunable with strain. Of the three devices, D3 has the highest gauge factor of 1513 (refer to Table 2), and the change in rectification ratio with strain is also high. The results indicate that there is a correlation between the cut-in voltages and rectification ratios. The device with the highest rectification ratio has a low cut-in voltage and vice-versa.

.

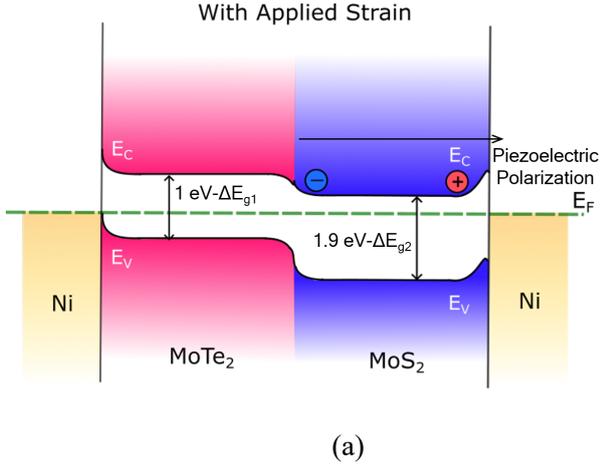 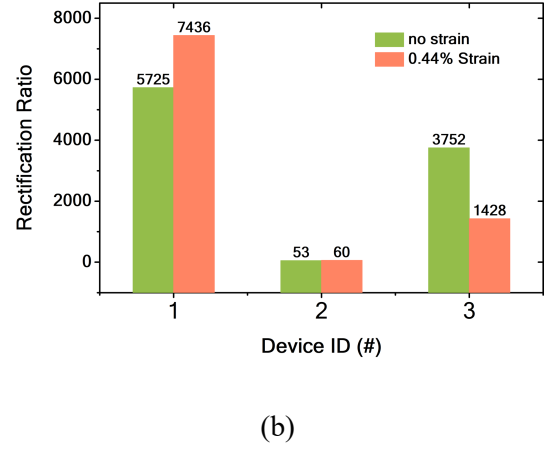

(a)            (b)

***Figure 4***: *(a) Band diagram of the 1L-MoS$_2$ and MoTe$_2$ junction under tensile strain, (b) Comparison of change in rectification for the devices with 0.44% applied strain*

The piezoelectric characteristics of these devices are analyzed by measuring open circuit voltage and short circuit currents with applied strain. Measurement schemes and error-mitigating strategies can be found in SI and are elaborated upon in Yarajena et al[29]. The open circuit voltages measured from these devices are shown in fig. 5(a-c). All three devices are strained with 0.44% with 4s ON and 4s OFF cycles. It can be observed that the device with the highest rectification ratio (D1) has the least open circuit voltage. In comparison, the device (D2) with the lowest rectification ratio has the highest measured open circuit voltages. For D2, $V_{OC}$ of more than 1V is observed at 0.44% applied strain. This value is significantly higher than any other piezoelectric device reported with a single 2D material. Device D3 has optimal open circuit voltages, rectification ratios and time constants and is therefore useful for both rectifying and piezoelectric applications.

$V_{OC}$ output with strain cycles has a time constant associated with it and is based on the device properties such as the resistance and the capacitances. We fit these exponentially decaying curves and extract the time constant for all three devices. The time constant of D2 is much higher (22 s), while that of D1 is very low (0.6 s) and is intermediate for device D3 (8s). The higher time constant observed for D2 indicates minimal screening, and we believe this is the result of the high cut-in voltage.

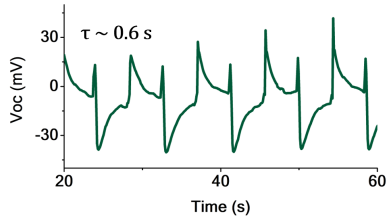 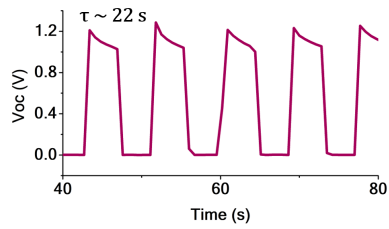 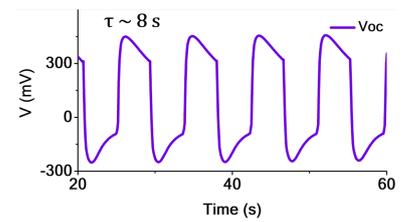

(a)            (b)            (c)

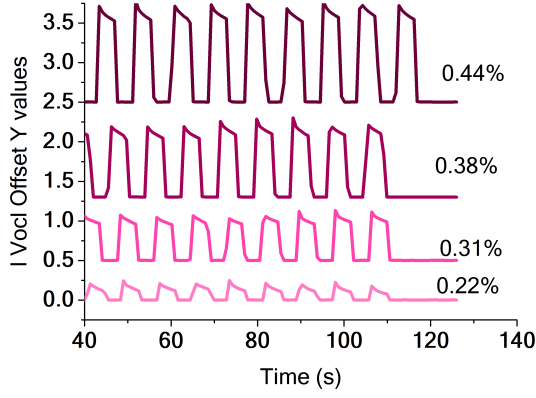 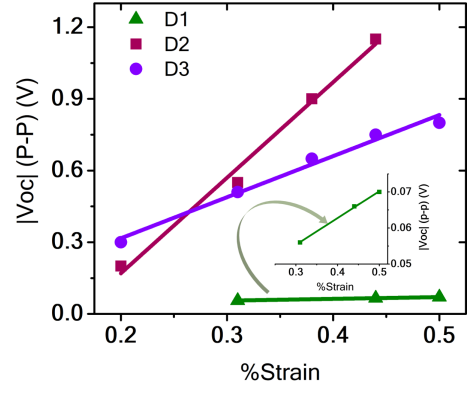

(d)            (e)

*Figure 5*: Open circuit voltages measured with 0.44% strain cycles for (a) D1, (b) D2 and (c)D3. Time constants at which voltage decays is labelled on each graph, (d) open circuit voltage of the D2 device changing with the applied strain, (e) plot showing open circuit voltages of devices D2, D3 with variable strain (in y-axis label, p-p refers to peak to peak value).

To further verify the effect of strain, open circuit voltages are measured at different strain levels. Figure 5(d) shows the open circuit voltages obtained for the D2 device at different strain values. With an increase in strain, the open circuit voltages increase. The plot of change in $V_{OC}$ with applied strain (fig. 5(e)) shows the linear dependency with applied strain for D1, D2 and D3 devices. Since device D1 has much lower open circuit voltages, the linear dependency is shown in the inset.

The short circuit currents follow a similar trend as the open circuit voltages (plots shown in SI). Device D2 has a high short circuit current, D1 has the least, while D3 has an intermediate value. However, the short circuit currents also depend on the rate of strain change. The piezotronic parameters obtained for devices D1, D2, and D3 are listed in Table 2. It is clear that there is a trade-off between the rectification ratios and piezoelectric voltage outputs for these devices. Higher the time constant of the device, the higher would be the energy-storing efficiency.

**Table 2**: List of the parameters obtained for D1, D2, and D3

|  | **D1** | **D2** | **D3** |
|---|---|---|---|
| Rectification Ratio (RR) | 5725 | 53 | 3752 |
| Maximum Gauge Factor (GF) at 0.44% strain | 1179 | 938 | 1513 |
| Open circuit Voltage ($V_{OC}$) at 0.44% strain (peak-peak) | 0.066 V | 1.15 V | 0.75 V |
| Peak short circuit current ($I_{SC}$) (pA) | 5 | 54 | 27 |
| Cut-in voltage (V) | 0.26 | 0.92 | 0.52 |
| Time constant from the open circuit voltages (s) | 0.6 | 22 | 8 |
| 2D device Area Approx. ($\mu m^2$) | 45 | 43 | 57 |
| Peak Power density ($mW/m^2$) | 3 | 690 | 177 |

Zhou et al. estimated a theoretical open circuit voltage of around ~350 mV at 0.53% strain for a $MoS_2$ device[12]. However, our devices, specifically a heterojunction device D2, surpassed this estimate with much higher open circuit voltages (>1 V) at 0.44% strain. Moreover, the device with the highest cut-in voltages (D2) has high open circuit voltages (>1V). Furthermore, the short circuit currents were observed to be higher for the device with a higher cut-in voltage (refer to SI), leading to increased energy output. Conversely, the device with lower cut-in voltages (D1) exhibited a reduced open circuit voltage of approximately 58 mV. On the other hand, device D3 with moderate cut-in voltage, exhibited optimal rectification ratios and piezoelectric outputs among these devices.

We believe that the cut-in voltages play an essential role in attaining high piezoelectric output. For devices where the charges get screened slowly, cut-in voltages are higher. This can also be verified from the time-constant analysis of open circuit voltage for various devices. An in-depth analysis of the transport in these heterostructures under piezoelectric behaviour is much more complex because of multiple junctions and van der Waal forces between the 2D layers. Detailed transport studies of these devices in future can unveil the exact reason for this behaviour. Based on the insights gained from this output characteristic analysis, it is evident that Schottky-pn hybrid diodes hold potential for the design of piezotronic rectifying junctions. The asymmetry in the junctions and finite cut-in voltage with the applied strain are the key factors to achieve this.

## Conclusion:

In conclusion, we have fabricated lateral heterojunctions of monolayer $MoS_2$ and multilayer $MoTe_2$ on a flexible substrate to study piezoresponse and rectification. The device architecture with a Schottky junction and a pn junction in series is devised to improve the piezoelectric response by reducing the charge screening effect. High piezoelectric outputs (>1V) were observed in these devices, which is the largest observed output voltage. This factor can have implications for energy harvesting devices and piezotronics devices. The device architecture can be further improved by optimizing the junction area and other characteristics of the junction. We also observe large rectification ratios reaching as high as ~5700, which will be useful for switching applications. The high switching ratio is observed in devices with large cut-in voltage devices, indicating a trade-off between the rectification ratio and the piezoelectric output.

## Experimental section

*Device Fabrication:* 2D material devices are fabricated on nano flex screen protectors[21]. $SiO_2$ base layer is deposited on the nano flex films using an RF magnetron sputtering tool from Tecport. The thickness of the deposited $SiO_2$ is around 90 nm with a standard deviation of 10 nm and is measured using a J.A. Woollam ellipsometer. 2D materials are obtained using standard mechanical exfoliation. Then the desired 2D material stack, i.e. few layer $MoTe_2$ and (1L) $MoS_2$, are stacked using the direct transfer method. Direct laser writer µpg Heidelberg is used for lithography to pattern photoresist (AZ5214E). Ni/Au (10/70 nm) metal films are deposited using an e-beam evaporation tool from Tecport.

*Characterization of devices:* Bending setup is developed in-house using a linear rail with a motor[21], 3D printed components to hold the device and PCBs. The entire setup is placed inside a Faraday cage to avoid electrostatic noise. B1500A Semiconductor device analyzer (SDA) from Keysight technologies is used for electrical measurements. Horiba LabRam HR is used to carry out Raman spectroscopy.

## Supporting information
Supporting information is available

## Conflict of Interest
The authors declare no conflict of interest

## Acknowledgements
We acknowledge funding support from MHRD, MeitY and DST Nano Mission through NNetRA

# Supplementary Information

# Engineering MoS$_2$-MoTe$_2$ Heterojunctions: Enhancing Piezoresponse and Rectification

Sai Saraswathi Yarajena[1], Akshay K. Naik[2]

1:saiyaras.95@gmail.com/sy785@cornell.edu, 2: anaik@iisc.ac.in
Centre for Nano Science and Engineering, Indian Institute of Science, Bangalore-560012, India.

## 1. Role of Screening on MoTe$_2$ Device

A few-layer MoTe$_2$ device with metal contacts (Ni/Au) is fabricated on a flexible substrate to check the strain-dependent characteristics. The I-V characteristics of this MoTe$_2$ device (Fig. S1(a)) indicate that the Ni contact with a few-layer MoTe$_2$ forms an ohmic contact. Ohmic contacts are not preferred for piezotronic devices because of the high screening rate. Fig. S1(b) shows the strain-dependent I-V characteristics of the few-layer MoTe$_2$ device. Since the device's resistance is low (in MΩ), the piezoelectric behaviour could not be observed in these devices.

While the MoTe$_2$ device exhibited near-ohmic behaviour with Ni contacts, the monolayer MoS$_2$ (Fig. 1(c)) formed a Schottky contact with Ni, as shown by the IV characteristics in Fig. 1(d). Furthermore, it was observed that the Schottky barrier height decreased with increasing tensile strain.

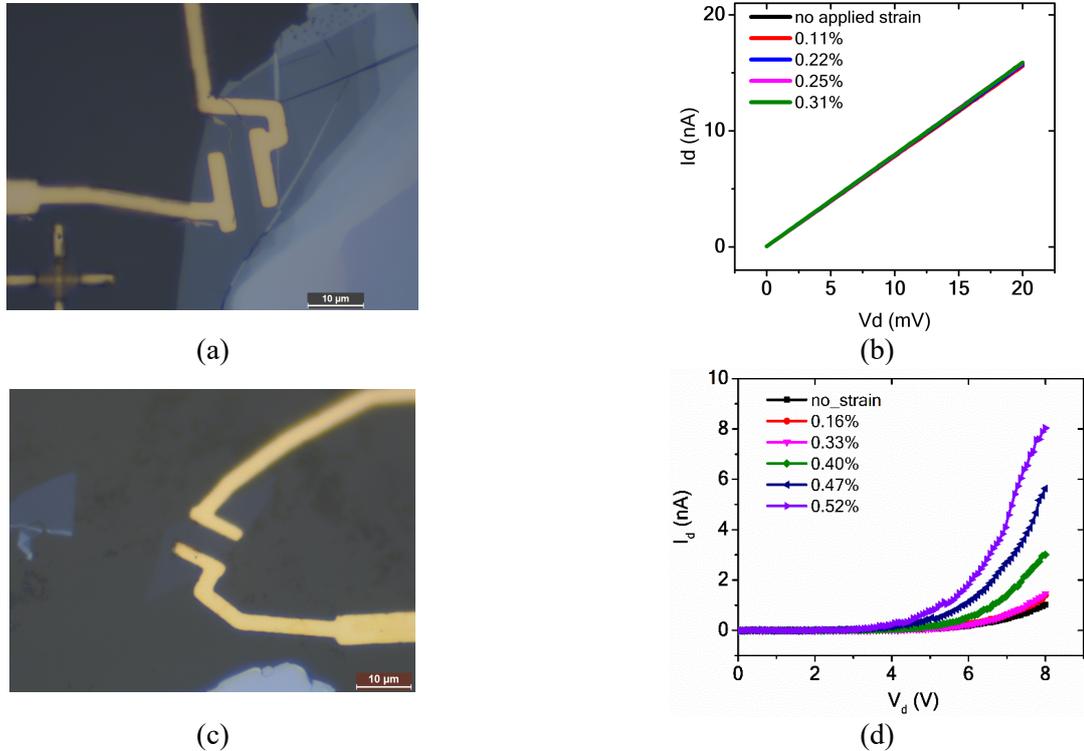

*Figure S1*: *(a, c) Optical Micrograph of few-layer MoTe$_2$ and monolayer MoS$_2$ devices, (b, d) strain-dependent I-V characteristics of few-layer MoTe$_2$ and monolayer MoS$_2$ devices respectively.*

## 2. Ideality factor estimation

The rectification ratio (RR) is calculated as

$$RR = \frac{I_{+V}}{I_{-V}}; \quad V > V_{cut-in} \qquad \text{Eq. (S1)}$$

The ideality factor of the rectifying junction devices is calculated from the slope (*m*) of the linear region in semi-log I-V characteristics, and it is given by

$$n = \frac{q}{m \ln(10) \, k_B T} \qquad \text{Eq. (S2)}$$

Where $q$ is the unit charge, $k_B$ is the Boltzmann constant, and $T$ is the temperature. $k_B T/q$ is also referred to as temperature-dependent voltage constant ($V_T$).

## 3. Rectifying ratio range

The rectification ratios of the six devices are shown in the table below. It has a wide range from 25 to 5725. Corresponding optical micrographs of all six heterojunction devices are shown in Fig. S2. The voltage at which these rectification ratios are estimated is mentioned in the x-axis labels. The ON-OFF current ratio is measured after the cut-in voltage.

| Device ID | Rectification ratio | Bias Voltage at which rectification ratio is calculated (V) |
|---|---|---|
| D1 | 5725 | 1 |
| D2 | 53 | 2 |
| D3 | 3752 | 2 |
| D4 | 25 | 3 |
| D5 | 823 | 5 |
| D6 | 602 | 4 |

In devices D2 and D4, the metal contact on $MoS_2$ touches the thicker $MoS_2$ and $MoTe_2$ layers adjacent to the 1L-$MoS_2$. Because of this, resistance in reverse bias reduces and rectification ratios for D2 and D4 could be low. Devices D5 and D6 have moderate rectification ratios because cracks are observed on the flakes after the transfer. Device D1 has the thickest $MoTe_2$ (~8 layers) of all the devices and, hence, the highest rectification among these devices. Other than geometric factors, interface traps also contribute to the contact resistance switch, which can affect the rectification ratio.

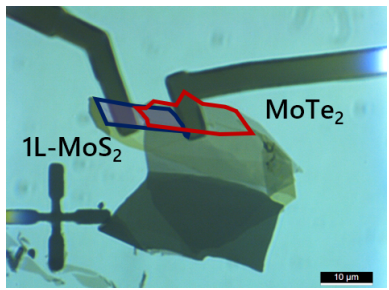 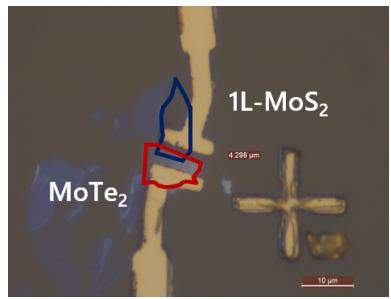 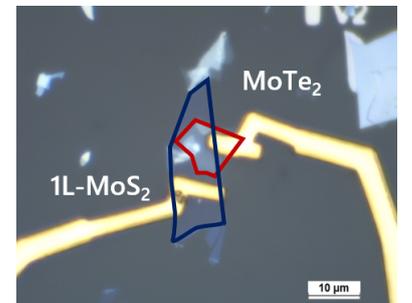

D1         D2         D3

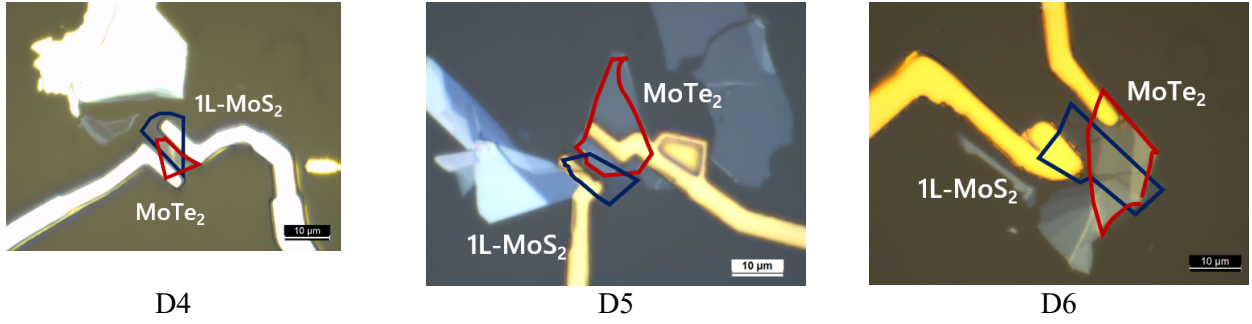

|  D4  |  D5  |  D6  |

**Figure S2**: *Optical micrographs of all six heterojunction devices (MoS₂ flake is outlined in blue and MoTe₂ flake in red)*

## 4. Gauge factor and Peak Power density

The gauge factor is estimated to check the sensitivity of piezoelectric/piezoresistive devices with applied strain. The current-based equation is used to compare the performance of these devices with that of literature. The gauge factor ($GF_I$) for the devices is calculated as

$$GF_I = \frac{[I(\varepsilon) - I(0)]/I(0)}{\varepsilon} \bigg|\text{constant } V \tag{S3}$$

Peak Power density is calculated as $(I_{SCmax} \cdot V_{OC\,max})/2$, where $I_{SC\,max}$ is peak short circuit current, and $V_{OC\,max}$ is peak open circuit voltage at applied strain ($\varepsilon$).

## 5. Measurement Setup

We have used a linear rail with a motorized stage to apply the bending strain, as shown in Fig. S3(a). 3D-printed flaps are used as support structures between the printed circuit board(PCB) and the motorized stage. Dual-sided PCBS are used to connect the connections. Flexible substrate is clamped between the two PCBs. The pitch of the metal electrodes on the PCB is designed to be the same as that of the electrodes on the PCB. Fig. S3(b) shows the image of the flexible substrate with applied bending strain. This is achieved by adjusting the distance between the two clamped ends. The whole setup is placed in a Faraday cage, and then the Triax cables are used to connect to the semiconductor device analyzer for electrical measurements.

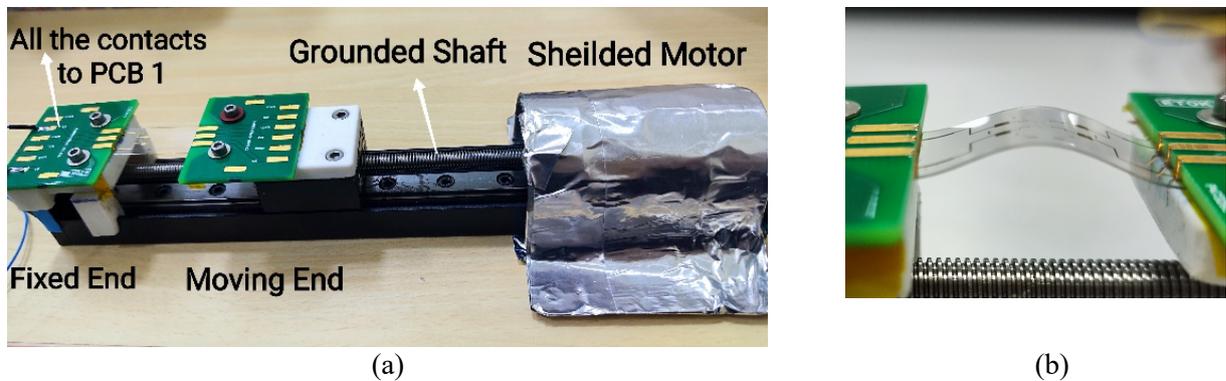

(a)          (b)

**Figure S3**: *Measurement setup showing various PCB connections, linear rail and a motor for measurements, (b) Image showing the applying bending strain on the flexible substrate*

## 6. Validating piezoelectric output

In the context of SC (short circuit) conditions, if the DUT (device under test) exhibits a net polarization attributed to the piezoelectric effect, there is an observed flow of charges through the device. Conversely, in the absence of piezoelectric polarization, the circuit ceases to meet short-circuit conditions. Consequently, no current flows through the device and any charges routed to the ground are from the contact electrification at the sample interface and electrostatics. In our current experimental configuration, depicted in Fig. S4(a), we employ two ammeters (or Source Measurement Units (SMU) with sub-picoampere resolution) to discern these phenomena. The currents, denoted as i1 and i2, are monitored by the two ammeters. It is imperative to ensure that the offsets of the ammeters are of comparable magnitude and polarity prior to conducting measurements. If there is a generation of current within the device, the currents i1 and i2 are anticipated to exhibit identical magnitudes but opposite polarities, consistent with Kirchoff's circuit laws. The net short circuit current ($i_{SC}$) in this scenario is determined by (i1-i2)/2. Hence, the piezoelectric output is validated by checking the direction of currents I1 and I2 from the short circuit current measurements. The I1 and I2 refer to the currents measured in differential configuration at each terminal. Fig. S4(b) shows the currents from two ammeter configurations for a heterostructure device. Detailed explanation on noise involved in the measurement setup can be found at Yarajena et al[1].

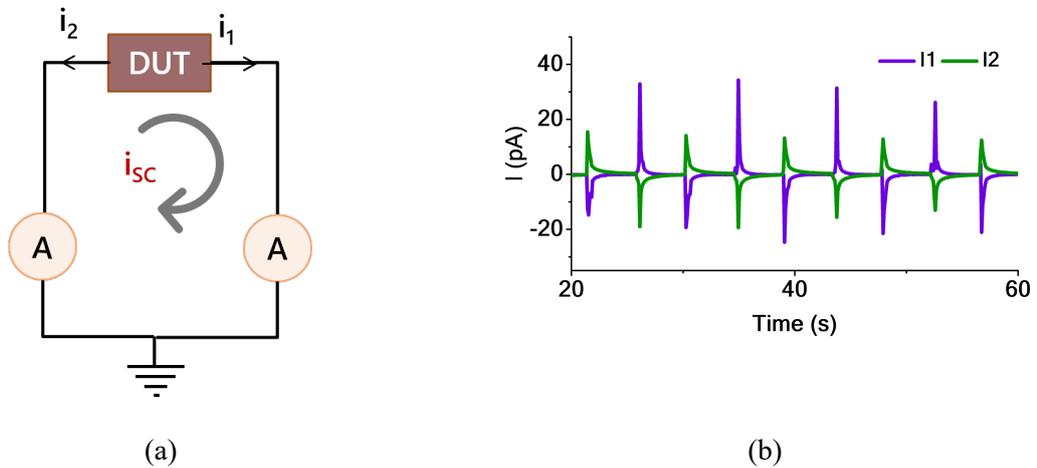

(a)  (b)

*Figure S4*: *The currents I1 and I2 measured under short circuit conditions from two ammeters.*

## 7. Short circuit currents

Short circuit current measurements are carried out on these devices, and Fig. S5 shows the short circuit currents ($i_{SC}=(i_1-i_2)/2$) measured for these devices. The time period and magnitude of the strain cycles are the maintained constant for all three measurements shown in Fig. S5(a-c) The short circuit currents also follow a similar trend as the open circuit voltages. Device D2 has a high short circuit current, D1 the least, while D3 has a moderate value.

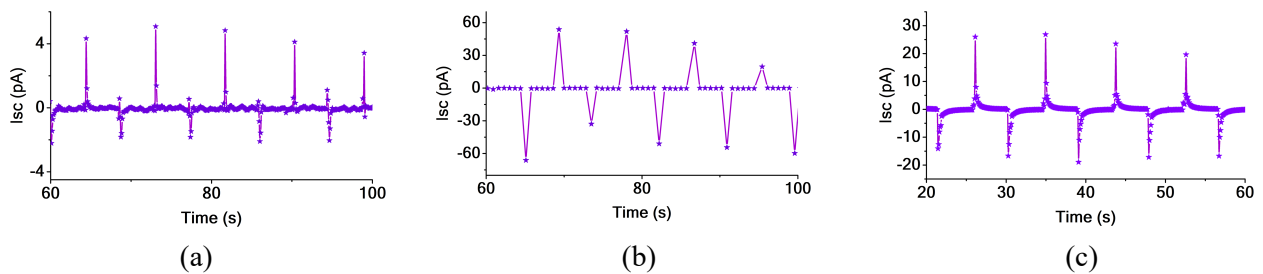

(a)  (b)  (c)

**Figure S5**: Short circuit currents measured with 0.44% strain cycles of (a) D1, (b) D2 and (c)D3

## 8. Variable time constant

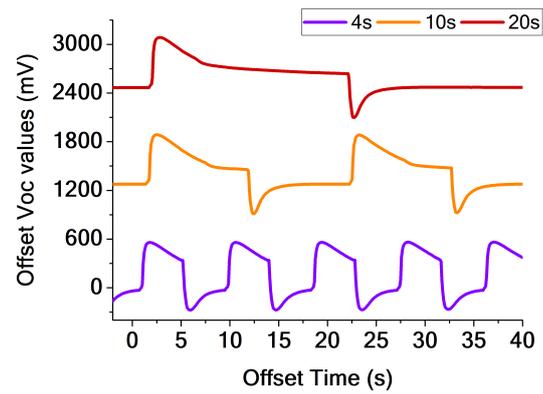

**Figure S6**: Dependence of open circuit voltages with time periods of applied strain.

Measurements are carried out for different time intervals to check the time-constant dependence. For device D3, the open circuit voltage measurement was carried out at 4s, 10s, and 20s intervals as shown in Fig. S6.

## References

[1]    S. S. Yarajena and A. K. Naik, *IEEE Transactions on Instrumentation and Measurement* **2024**, DOI 10.1109/TIM.2024.3500064